\documentclass{article}
\usepackage{graphicx} 
\usepackage{epsfig}     
\usepackage{graphicx}   
\usepackage{url}
\usepackage{natbib}     
\usepackage{babel}       
\usepackage{amsmath}
\usepackage{subcaption}

\date{January 2025}

\begin{document}

\title{Origin likelihood functions for extreme-energy cosmic rays \\[1ex] \large Submitted for the proceedings of ECRS-2024}

\author{Leonel Morejon$^{1}$
\vspace{2mm}\\
\it $^1$Bergische Universit\"{a}t Wuppertal,\\ 
\it Gausstrasse 20, 42117 Wuppertal, Germany
}

\maketitle

\begin{abstract}
Unlike neutrinos and photons arriving from extra-galactic sources, ultra-high energy cosmic rays (UHECRs) do not trace back to their origins due to propagation effects such as magnetic deflections and energy losses. For ankle energies, UHECRs can propagate for hundreds of megaparsecs with
negligible energy losses but the directional information is lost after a few megaparsecs. On the other hand, at the highest energies the directions are kept for larger distances due to the increased rigidity but the interaction rates with the cosmic microwave background strongly suppress the
cosmic rays within a few to tens of megaparsecs. Therefore, UHECRs of extreme energies, such as the Amaterasu event recently reported by Telescope Array, are of particular interest to identify the sources within our galactic neighborhood. However, photonuclear interactions are stochastic in nature and produce changes in the nuclear species emitted, which difficults the task of estimating
the likelihood distribution of its origin. This work discusses a novel procedure to estimate the likelihood of the origin for extreme-energy cosmic rays based on probability distributions for UHECR stochastic interactions. The method is applied to the Amaterasu event and compared to recently published works which employ Monte Carlo codes (e.g. CRPropa) in their analysis. The advantages of the method presented here are demonstrated by the increased resolution and the ease of computation unlike other approaches employed so far. The results presented indicate that the localization of the origin of extreme energy cosmic rays could be possible in some cases without knowledge of the original composition.
\end{abstract}

\section{Introduction}

The existence of cosmic rays with extreme energies ($E > 10^{20}$ eV) was already well established 20 years ago \citep{RevModPhys.72.689}, and it was clear they hold scientific interest regarding their origin although their nature was not known \citep{Cronin2005}. Since then, we have become increasingly confident that most cosmic rays at the highest energies must be heavier than protons \citep{PierreAuger:2024flk}. 

The large density of cosmic microwave background (CMB) photons renders the universe opaque for extreme-energy cosmic rays (ExECR) originating beyond a few tens of megaparsecs from Earth, because of different interactions causing strong energy losses. For ExECR protons, photopion production is the dominant interaction, while nuclei photodisintegrate as a result of the excitation of the giant dipole resonance. The expected horizons (Greisen-Zatsepin-Kuz'min horizon for protons and a photodisintegration horizon for nuclei \citep{1976ApJ...205..638P}) reduce considerably the volume of possible origin. Additionally, with the large rigidities of ExECR, the influence of extragalactic magnetic fields is negligible and the arrival directions correlate strongly with the origin, provided that the coherent component of the galactic magnetic field is well understood (see recently updated models \citep{Unger_2024a,korochkin2024coherentmagneticfieldmilky}).

These reasons argue for the potential of ExECRs to identify the sources in our cosmic vicinity, but the reduced flux of these events (about one per square kilometer per century) requires an event by event analysis. Furthermore, photodisintegration interactions of ExECR nuclei are stochastic in nature and require a probabilistic approach for a suitable description. In the case of the recent Amaterasu event reported by Telescope Array \citep{doi:10.1126/science.abo5095} the likely origin was estimated with the help of the Monte Carlo code CRPropa \citep{crpropa3.2} to include the stochastic interactions and account for the range of possible species at Earth \citep{Unger2024}. However, there are a number of limitations of Monte Carlo methods which can be overcome if the sought outcomes have known distributions. This work discusses a novel procedure to estimate the likelihood of the origin for extreme-energy cosmic rays based on probability distributions for UHECR stochastic interactions. The method is applied to the Amaterasu event and compared to the recent analysis employing CRPropa \citep{Unger2024}. In contrast to other probabilistic approaches, such as the one employed in \citet{bourriche2024localvoidcomprehensiveview}, the distributions used here \citep{Morejon:20239X} are computed from interaction rates and the theory of matrix exponential distributions, thus requiring no Monte Carlo simulations.


\section{Origin likelihood functions}

The mean interaction length of ExECRs depends on the cross section for the relevant interaction and the density of the target background photons. It is given by the inverse of the interaction rate $\lambda(\gamma)$ per unit distance 

\begin{equation}
    \lambda(\gamma) = \frac{1}{2\gamma^2}\int_0^{\infty} \frac{n(\epsilon)}{\epsilon^2} d\epsilon \int_0^{2\epsilon \gamma} \varepsilon \sigma (\varepsilon) d\varepsilon
\label{eq:interaction_rate}
\end{equation}

\begin{figure}[t]
 \centering
  \includegraphics[scale=0.6]{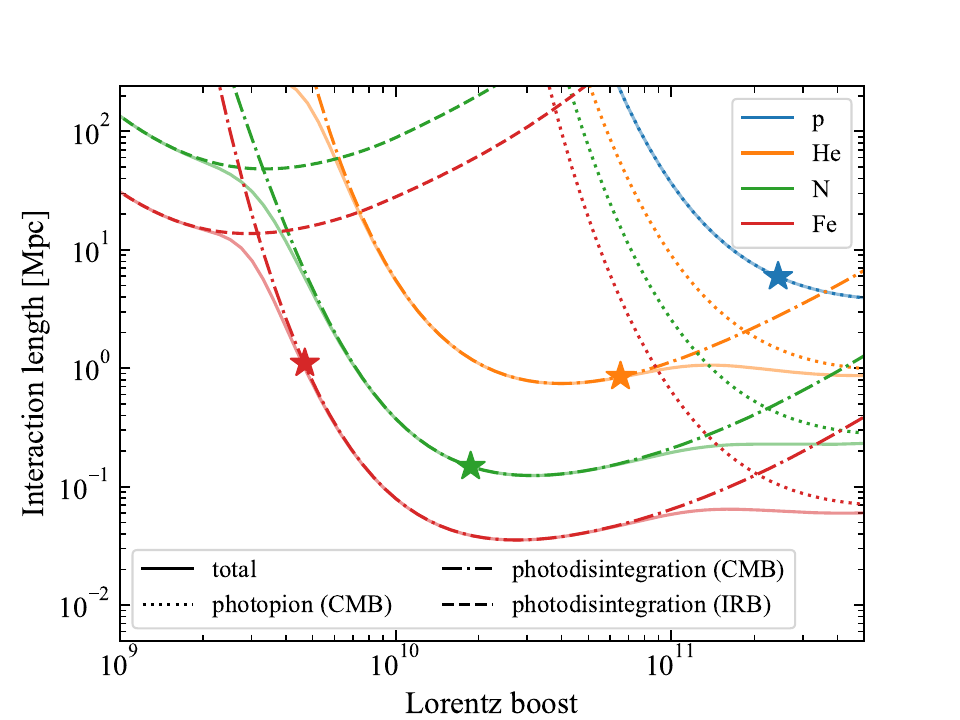}
 \caption{Energy loss lengths versus Lorentz boost for different nuclear species. The species are indicated by colors and the interactions by different line styles. Stars indicate interaction lengths and Lorentz boost for the corresponding species having the energy of Amaterasu (244 EeV).}
 \label{fig:interaction_lengths}
\end{figure} 

Figure~\ref{fig:phasespaceiron} shows the interaction lengths for different nuclei (indicated by the colors) in the range of Lorentz boosts relevant for ExECRs. The stars denote the values of boost and interaction length if the Amaterasu event was of the species indicated by the color. For protons (or neutrons) photopion interactions with the CMB (dotted, blue) are the dominant interaction, while for nuclei, photodisintegration interactions with the CMB (dash-dotted) dominate over interactions with the infrared background (IRB, dashed lines) over most of the boost range. Pair production interactions are also possible but the corresponding interaction lengths are subdominant in this energy range and their effect can be neglected, especially for ExECRs originating within $\sim 100$ Mpc. We will focus here in the hypothesis that Amaterasu is not a proton but a heavier nucleus, following the argument that fits of spectrum and composition indicate heavier nuclei at the maximal energies and a maximal rigidity of $R_{\mathrm{max}} \in [3, 5]$ EV \citep{Unger_2024}. For such extreme rigidities, and the reported energy of Amaterasu, nuclei accelerated at the source could potentially have a mass $A \approx E_0/R_{\mathrm{max}}/\eta \in [106, 176]$, with $\eta = Z / A \simeq 0.46$. The boost conservation characteristic in photodisintegration interactions implies that, as ExECR disintegrate, the products preserve roughly the rigidity since $\gamma = E_0/A=\eta R$ with $\eta$ changing slightly from species to species. This justifies conceiving the propagation of the ExECR from source to Earth as ballistic, since deflections are only significant once they enter the galaxy. Thus, the main quantity that needs to be understood is the change of nuclear species due to successive photodisintegration interactions, with the conservation of the boost. The deflections inside the galaxy can be studied with CRPropa as in other works \citep{doi:10.1126/science.abo5095,Unger2024,bourriche2024localvoidcomprehensiveview} but due to the small distances involved ($\sim 15$ kpc) the probability of interaction is negligible. Therefore, the composition should remain unchanged from entering the galaxy.

The stochastic disintegrations of nuclei lead to the production of multiple secondaries with different probabilities, resulting in complex disintegration chains that require a probabilistic description. This process is a type of continuous-time Markov chain and is described by matrix exponential functions \citep{Morejon:20239X}. This description was motivated by regularities observed in the photodisintegrations of ExECR nuclei of masses up to A=208 \citep{Morejon2020} and is able to explain them. The probability density for the distance needed to cascade from the initial species $S_0 =(Z_0, A_0)$ to a set of possible ending species $\{S_1^f, S_2^f, ..., S_k^f\}$ (with $S^f_n = (Z^f_n, A^f_n)$) is 

\begin{equation}
    f(L)= \boldsymbol{\alpha_0} \exp \left( \mathbf{\Lambda} L\right) \mathbf{\Lambda} \mathbf{e}
    \label{eq:pdf}
\end{equation}

where $\boldsymbol{\alpha_0}$ is a vector of zeros except for the first species $S_0$ where it is one, and expresses the initial fractions for each species. 
$\boldsymbol{\Lambda}=\boldsymbol{\Lambda}(\gamma)$ is a matrix formed by all the interaction rates of all species that are involved in the transition from the initial to the final ones, evaluated at the boost of interest $\gamma$

\begin{figure}
    \centering
    \begin{subfigure}[b]{0.49\textwidth}
        \centering
         \includegraphics[scale=0.5]{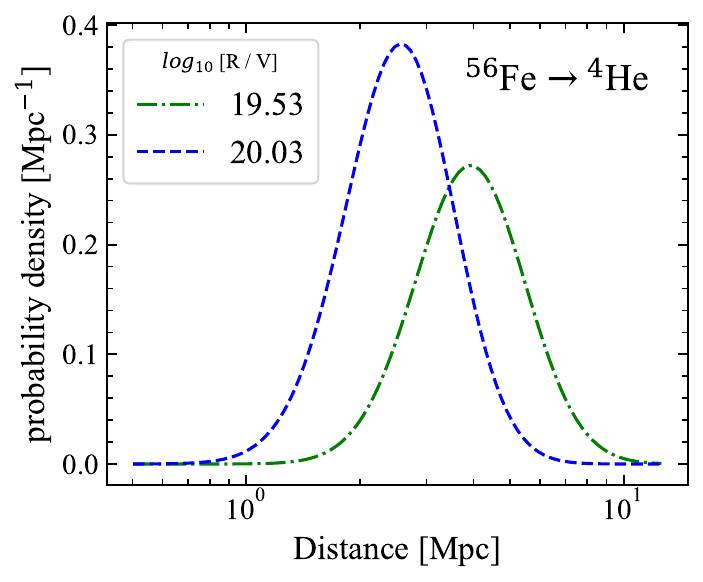}
    \end{subfigure}
    \hfill
    \begin{subfigure}[b]{0.49\textwidth}
        \centering
        \includegraphics[scale=.435]{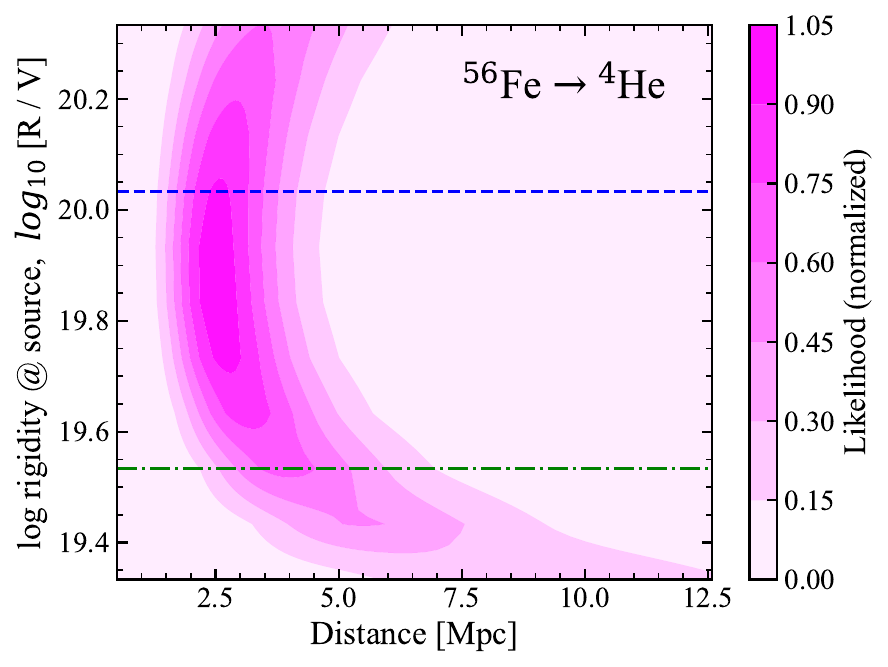}
    \end{subfigure}
    
     \caption{Left: Probability distributions with distance for the disintegration of $^{56}$Fe until the remnant is $^4$He. Two rigidities indicated by the color and line style are shown in both figures. Right: likelihood distribution as a function of the rigidity at the source and the origin of the parent.}
    \label{fig:example_density_distributions}
\end{figure}

\begin{equation}
    \mathbf{\Lambda}(\gamma) = 
    \begin{pmatrix}
        -\lambda^{\rm{tot}}_{S_0} & \lambda_{S_0 \to S_1} & \lambda_{S_0 \to S_2} & \lambda_{S_0 \to S_3}& ... & \lambda_{S_0 \to S^f}\\
        0 & -\lambda^{\rm{tot}}_{S_1} & \lambda_{S_1 \to S_2} & \lambda_{S_1 \to S_3} & ... & \lambda_{S_1 \to S^f}\\
        0 & 0 & -\lambda^{\rm{tot}}_{S_2} & \lambda_{S_2 \to S_3} & ... & \lambda_{S_2 \to S^f}\\
        0 & 0 & 0 & -\lambda^{\rm{tot}}_{S_3} & ... & \lambda_{S_3 \to S^f}\\
        ... & ... & ... & ... & ... & ... &\\
        0& 0& 0& 0& ... & -\lambda^{\rm{tot}}_{S^f}\\
    \end{pmatrix}
\end{equation}

\noindent with $\lambda_{S_n \to S^f} = \sum_{j=1}^k \lambda_{S_n \to S^f_j}$ is the sum of all transition rates leading to nucleus in the set of ending species and $\lambda^{\mathrm{tot}}_{S_n} = \sum_j \lambda_{S_n \to S_j}$ . Examples of these distributions are shown in Figure~\ref{fig:example_density_distributions} (left) for a starting species $^{56}$Fe and ending species $^4$He. Varying the boost yields a two dimensional likelihood function as shown in Figure~\ref{fig:example_density_distributions} (right) where the rigidity is obtained from the boost according to the previous relation. This method works for all final species that are produced only once in the whole cascade, which is the case of all products except the lightest (e.g. protons and neutrons) because they can be produced in multiple points along the cascade and not only as a final state. For such particles the total yield as a function of distance is also matrix-exponential distributed and the probability density can be obtained through a transformation by rewards of Equation~\ref{eq:pdf} \citep{Bladt2017} where the "rewards" are the number of protons produced in each of the possible transitions $\lambda_{S_i \to S_j}$. The likelihood function for these products is the convolution of the distribution obtained from the rewards transform and a flat function of the distance denoting a homogeneous emission throughout the distance range considered.

The complete set of origin likelihood functions can be obtained by changing the set of final nuclear species. Figure~\ref{fig:phasespaceiron} shows these functions where the final sets were chosen to be comparable to the simulation results in Figure~1 of \cite{Unger2024}. Note that the functions are restricted in rigidity by the constraint that the energies of the observed nuclei are distributed according to a gaussian with mean value $E_{\mathrm{low}}=1.64 \times 10^{20}$ eV and a standard deviation corresponding to the measurement uncertainty $\hat \sigma = 0.19 \times 10^{20}$ eV \citep{Unger2024}. An important distinction is that the transparency in Figure~\ref{fig:phasespaceiron} reflects the relative likelihood of observed nuclei having originated from the corresponding distance and rigidity. However, the transparency of scattered points in Figure~1 of \cite{Unger2024} only reflects the gaussian density for the deviation from the mean energy. This is an advantage of the likelihood functions obtained by the method presented here with respect to Monte Carlo simulations: the relative probabilities are well determined with arbitrary desired precision. This is more difficult with Monte Carlo methods because the regions of the phase space are only well constrained when a sufficient number of events have been observed. Furthermore, areas of the phase space with no observations cannot be constrained at all, and the complete space of possible events is inaccessible.

\begin{figure}[t]
 \centering
  \includegraphics[scale=0.65]{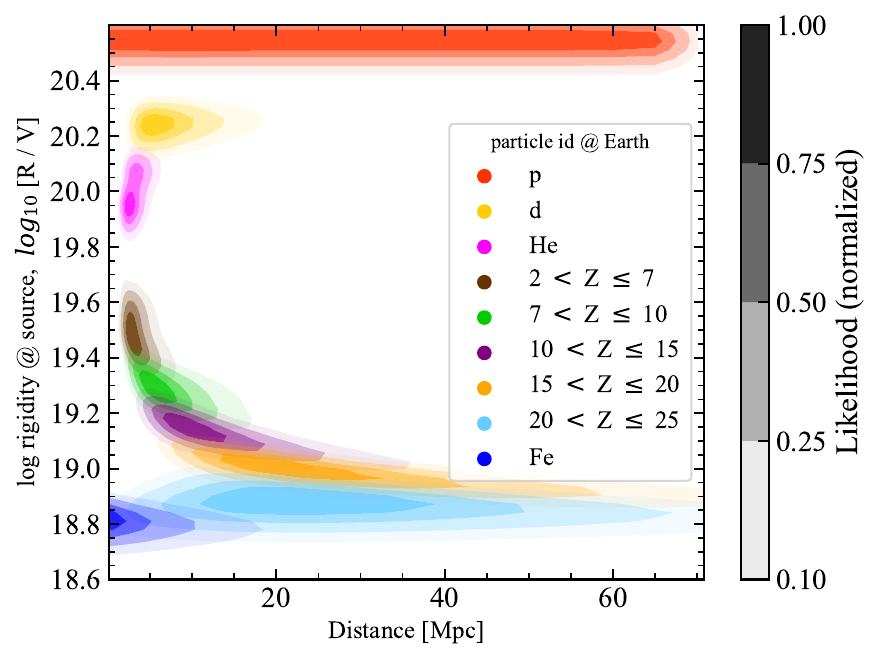}
 \caption{Origin likelihood distribution for the injection of iron. }
 \label{fig:phasespaceiron}
\end{figure} 

\begin{figure}
 \centering
  \includegraphics[scale=0.65]{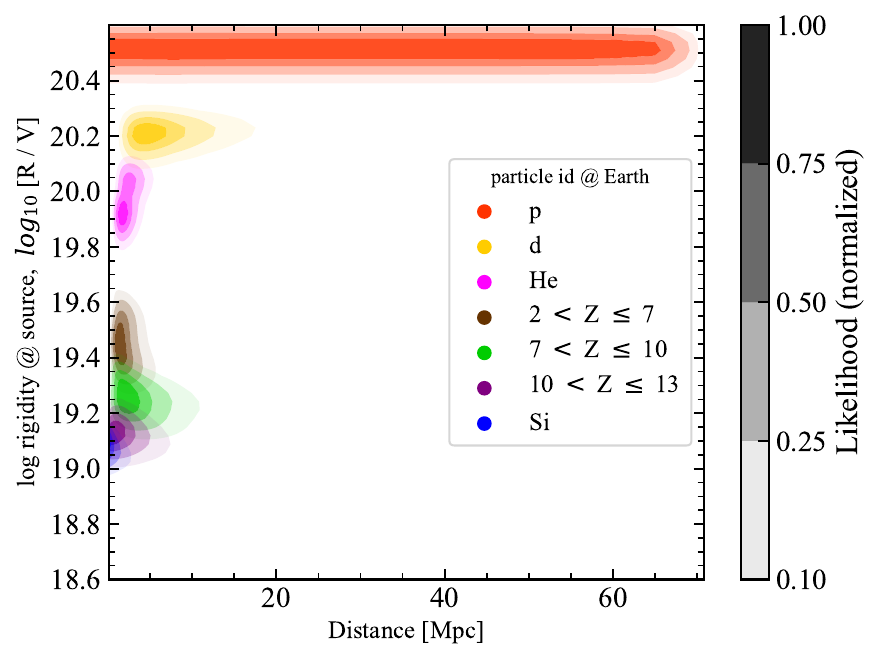}
 \caption{Origin likelihood functions for the injection of silicon.}
 \label{fig:phasespacesilicon}
\end{figure}

\begin{figure}
 \centering
  \includegraphics[scale=0.65]{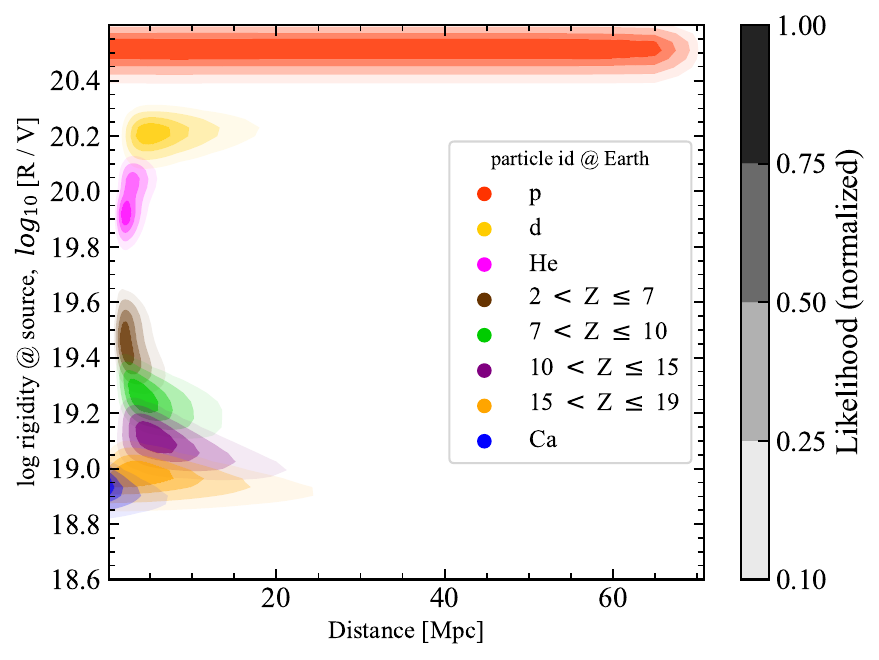}
 \caption{Origin likelihood functions for the injection of calcium.}
 \label{fig:phasespacecalcium}
\end{figure}

Figure~\ref{fig:phasespaceiron} can be understood from the shape of the likelihood function in Figure~\ref{fig:example_density_distributions} (right) and the interaction lengths in Figure~\ref{fig:interaction_lengths}. The interaction lengths are minimal at boosts of $2-3 \times 10^{10}$ for most nuclear species, and the cascades happen in a shorter range since, due to boost conservation, all products have also minimal interaction lengths. Products in this boost range result after shorter propagation lengths, and as the boosts deviate from these values the needed distances for ending in the same products increase. This is observed in Figure~\ref{fig:example_density_distributions} (right) where the distances are shorter for a small range of rigidities and increase outside of this range. Since the likelihoods for the observed species are restricted to within a range of the observed energy, the rigidity values are larger the lower the mass of the products $R_{\mathrm{obs}}=E_{\mathrm{obs}} / A_{\mathrm{obs}} / \eta$. This also explains how, for lower masses of the observed species, the distributions are more narrowly localized in distance, but for larger masses they are more spread since they fall in a range of boosts where the interaction lengths are larger. The spread in rigidity observed in sets with multiple final species results from the overlapping gaussian distribution in energy for the different species. For example, the group indicated as He (magenta) shows a two-lobed shape corresponding to the two species included $^3$He and $^4$He. The separation between species is less clear for larger sets because they fall closer in rigidity. For the case of deuterium (yellow) only one final species is included and the range of rigidities is narrowly defined (as expected for a single species) which is also observed for the starting species (blue). For protons, we also observe this a similar rigidity range although some narrowing with distance is to be expected, as protons produced at larger distances and close to the lower edge of rigidity would lose some energy due to photopion losses and fall outside of the minimal rigidity. However, this effect is expected to be reduced and was not included in the results shown in here (Figures~\ref{fig:phasespaceiron}-\ref{fig:phasespacecalcium}). 

These observations are verified in the origin likelihood functions for other starting species, shown in Figures~\ref{fig:phasespacesilicon}-\ref{fig:phasespacecalcium} where $^{28}$Si and $^{40}$Ca have been used respectively. These starting species have lower masses than $^{56}$Fe and correspondingly can only be observed at Amaterasu energies for larger values of the boost, which explains how the origin likelihood for their observation (blue) presents a reduced distance range. At the same time, the likelihood functions for the observed lower masses fall in the same ranges of rigidities and distances compared to the case of starting $^{56}$Fe. 

These results underline the importance of developing experimental methods to determine the composition on an event-by-event basis which is applicable to ExECRs. The likelihood distributions shown can be very constrained in distance for certain mass groups, thus a determination of the composition of an event could help establish its origin to within a few megaparsecs, reducing the list of possible origins to possibly a specific galaxy. This conclusion is independent of the original composition: for example, the origin of events detected in the charge range $2 < Z \leq 7$  is remarkably similar for the three different starting species considered $^{28}$Si, $^{40}$Ca and $^{56}$Fe (Figures~\ref{fig:phasespaceiron}-\ref{fig:phasespacecalcium}). 

\section{Summary}
The importance of extreme-energy cosmic rays ($E>10^{20}$ eV) for uncovering the closest sources is discussed. Their short interaction lengths and large rigidities are assets in restricting the distance of origin to a few megaparsecs. The origin likelihood functions have been computed based on the probability distributions for cascades of ExECR in extragalactic propagation. Different starting species were considered ($^{28}$Si, $^{40}$Ca and $^{56}$Fe) showing that the likely origin distributions are independent of the original composition if the original species have disintegrated sufficiently and for some range of ending species. The importance of a method to determining the composition event-by-event at the highest energies could be the corner stone in deciphering the origin of UHECRs. A good understanding of the galactic magnetic fields will also be required, but the current uncertainties might be sufficient for sources within $\sim 10$ Mpc.

\section*{Acknowledgements} 
This work has received support from the DFG under grant number 445990517 (KA 710). I am very thankful to the organizers of the ECRS-2024 for an enriching and productive conference. 

\bibliographystyle{apalike}
\bibliography{main}

\end{document}